# Biomimetic Ultra-Broadband Perfect Absorbers Optimised with Reinforcement Learning

Trevon Badloe†, Inki Kim†, and Junsuk Rho*, †, ‡,

†Department of Mechanical Engineering, Pohang University of Science and Technology (POSTECH), Pohang 37673, Republic of Korea

‡Department of Chemical Engineering, Pohang University of Science and Technology (POSTECH), Pohang 37673, Republic of Korea

*E-mail: jsrho@postech.ac.kr

**Abstract**

By learning the optimal policy with a double deep Q-learning network (DDQN), we design ultra-broadband, biomimetic, perfect absorbers with various materials, based the structure of a moth's eye. All absorbers achieve over 90% average absorption from 400 to 1,600 nm. By training a DDQN with moth-eye structures made up of chromium, we transfer the learned knowledge to other, similar materials to quickly and efficiently find the optimal parameters from the ~1 billion possible options. The knowledge learned from previous optimisations helps the network to find the best solution for a new material in fewer steps, dramatically increasing the efficiency of finding designs with ultra-broadband absorption.

**Introduction**

Metamaterials that exhibit a wide range of functionalities have been designed over the past few decades [1-6]. Of these metamaterials, perfect absorbers offer numerous benefits over conventional absorbers such as increased effectiveness, miniaturisation and the ability to be designed for a specific narrow or broadband range [7-10]. There are a variety of methods for achieving perfect absorption in metamaterials, such as 1D and 2D gratings [11,12], Fabry-Perot resonators [13,14] and structured metasurfaces [15,16]. There are a variety of structures that have been used for broadband absorption, such as crosses, cones, and simple square and circular rods [17-21]. These metamaterial absorbers can be used in applications that range from sensors, camouflage and wireless communication, to solar photovoltaics and thermophotovoltaics. Here, we take inspiration from nature in the form of moths' eyes to design an almost perfect, ultra-broadband absorber that spans across the entire visible and near infrared range (from 400 to 1,600 nm).

Antireflective structures in the eyes of moths have dimensions smaller than the incident wavelength of light [22]. These act as regions of graded refractive index between the interface of the ambient medium, i.e. air, and the surface of the eye. This allows moths to have almost no light reflected from their eyes, which helps them to avoid predators and survive for longer. This kind of biomimetic structure has been utilised in applications such as smart phone antireflective coatings and increasing solar cell efficiency [23,24].

The moth-eye structure acts as an impedance-matching mechanism that has a smooth transition between one medium to the other as shown in Figure 1(d). In this way, the light never reaches a point where there is a sudden change in refractive index, and subsequently there is almost no reflected light. Moth-eye shaped structures that are arranged in a hexagonal lattice have been studied to have a better performance than in a square lattice [25], so we naturally choose to use a hexagonal lattice here. By adding a metallic reflector to the bottom of our device to act as a mirror, the transmission is forced to be zero, while the antireflective layer simultaneously limits the reflection. By using a dielectric spacer layer between the moth-eye structures and the reflective back layer, we create a two dielectric-metal interfaces where surface plasmons can be excited and can interact, leading us to almost perfect absorption from 400 to 1,600 nm with various metallic materials.

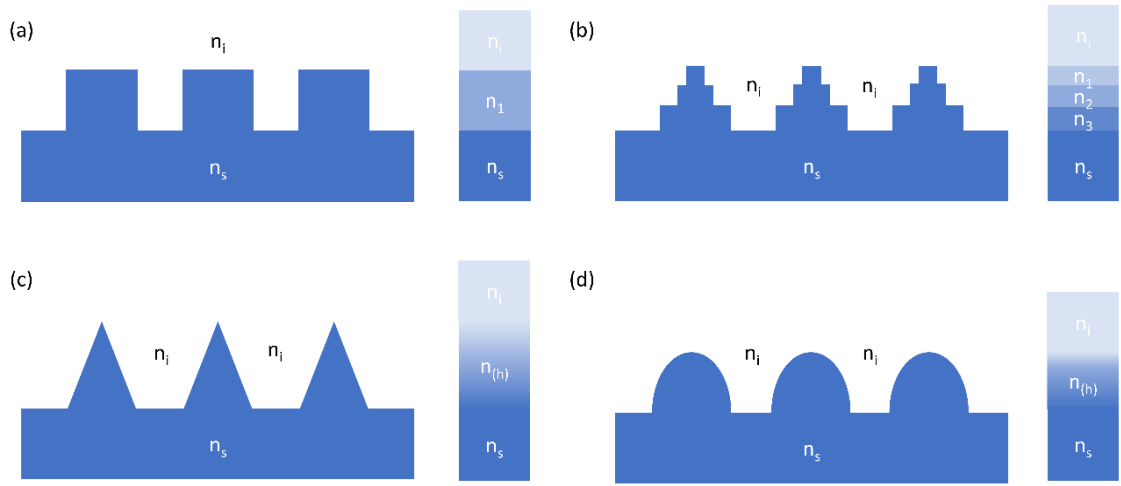

**Figure 1**. Effective refractive index of different shaped structures. a) An abrupt square step, b) tapered square steps, c) a pyramid and d) moth-eye shape. Moth-eye structures have a parabolic surface, which makes the effective refractive index change gradually, allowing the incoming light to never experience a sudden change in refractive index which would cause reflections. Where $n_x$ is the refractive index of each material.

Moth-eye structures are parabolic in shape, so to understand the mechanisms behind the absorption, we investigate the effective refractive index of the structures. The definition of the parabolic surface is given by $S = \frac{cr^2}{1+\sqrt{1-(\kappa+1)c^2r^2}}$. Where a conic constant ($\kappa$) of -1 gives a parabolic shape and the values c = 1/R, and $r^2 = u^2 + v^2$ are the inverse of the radius of curvature and the position on the surface respectively. The effective refractive index can then be understood as a function of the z-position of the height of the moth-eye structure, described by:

$$n_{eff}(h) = n_{moth\ eye} \times \frac{A_{moth\ eye}}{A_{base}} + n_{air} \times \frac{A_{base} - A_{moth\ eye}(h)}{A_{base}}, \quad (1)$$

Where $A_{moth\ eye}$ is the area of the moth-eye structure as a function of its height, $A_{base}$ is the area of the base of the shape, and $n_{moth\ eye}$ and $n_{air}$ are the refractive index of the material and the surrounding media, here air, respectively. Therefore, for a parabolic shape, the effective refractive index can be given by:

$$n_{eff}(h) = \frac{\pi}{2\sqrt{3}} \times \left(1 - \frac{h}{h_0}\right) \times \frac{r_{moth\ eye}}{r_i} \times (n_{moth\ eye} - 1) + 1, \tag{2}$$

Where $h_0$ is the height of the moth-eye structure, $r_{moth\ eye}$ is the radius of the structure and $r_i$ is the radius of the circle that can fit inside the hexagonal lattice unit cell.

Figures 1(a) and (b) show discrete steps in the refractive index, where the incoming light will meet a definite and abrupt change and will therefore create reflections. The pyramid structure in Figure 1(c) has a constant gradient in the refractive index, which is analogous to the steps in Figure 1(b), but with much finer steps. This helps to limit the reflections between each layer but is not as effective as the moth-eye structure. Comparisons can be found in the supplementary information.

Artificial neural networks have been used in nanophotonics for estimating the optical response, optimising the design parameters and the inverse design of metamaterials [26-29]. We use an unsupervised learning algorithm called double deep Q learning, to optimise the parameters of ultra-broadband, moth-eye structure perfect absorbers for several different metallic materials. By exploring and exploiting important regions of the parameter space, the model quickly optimises the absorption of the structure and chooses the appropriate materials for the substrate and spacer layers to find the highest average absorption over the specified wavelength range. We targeted over 90% average absorption over the whole region.

Since neural networks learn the general relationships between the input and output values by optimising the weights and biases of each node, they are extremely useful in optimising similar structures in similar regimes. As the network learns the general relationships between each parameter, after training, the same network with the same weights and biases can be used to optimise the parameters for a completely new material, even though the optical properties of the material are completely different. This concept, known as transfer learning, is often used in machine learning and has also been shown to be effective in nanophotonics for helping a network to learn about similar problems [30]. If we were to do a parameter sweep over all the possible options presented here, we would have around 1 billion possible different combinations for the device. After optimising the dimensions and spacer and substrate materials for one metal, the information from all of the previous simulations leading to the result are almost useless if we change the material used for the moth eye structure. Transfer learning allows us to reuse what the network learned for one material and use that as a starting point for optimising another. This helps the new agent to learn the best policy for the new regime with a new material, which greatly speeds up the learning and thereby the optimisation process.

**Results and Discussion**

The reinforcement learning code was written in Python using the PyTorch framework and the simulations were all performed using the commercially available FDTD solver, Lumerical FDTD Solutions. All material fits were checked and edited to ensure that they were as accurate as possible. Monitors were placed in the substrate, and behind the source to measure the transmitted and reflected responses respectively. The absorption was then calculated as,

$$A = 1 - T - R, \tag{3}$$

With care taken with regards to the sign of the response of the monitor, since the recorded response is negative from the transmission monitor. Anti-symmetric boundary conditions were used on the x-axis, with symmetric boundary conditions on the y-axis due to the symmetry of the design to help speed up simulations. Boundary conditions on the z-axis were set to perfectly matched layers (PML), with mesh settings chosen to be adaptive depending on the size of the given structure while ensuring that the simulation time was fast enough with no decrease in accuracy. A periodic Bloch plane wave in the range 400 to 1,600 nm was used to illuminate the device from above (i.e. in the negative z-direction). The metallic back reflector was taken to be infinite at the bottom of the device.

As always with reinforcement learning in nanophotonics, the limiting factor on the speed of learning is from the simulations themselves. If we choose to have a larger mesh, we gain some speed but lose a lot of accuracy. This hinderance also applies to parameter searches by human researchers, so using intelligent searches where the learned information can be reused in a second environment to speed up subsequent optimisations is a great benefit of this method.

**Reinforcement learning setup**

As described in previous work [31], DDQNs can be used to optimise the parameters of a device for a specific optical response. Here, we optimised the broadband perfect absorber to have over 90% absorption over the visible and near infrared spectrum, from 400 to 1,600 nm, by first learning the best policy for moth-eye structures made up of chromium (Cr) and then transferring the knowledge to quickly and efficiently optimise the parameters for other metals. The DDQN is made up of two models, a target network and a policy network. The policy network is trained after each action on a minibatch of pervious results, and the target network is used to predict the Q-values for the given state. An experience replay memory of 10,000 state, action, reward, next action transitions are initialised to keep track of what the network has seen, to be sampled from for each minibatch while learning. To train the policy network, a minibatch sample of 32 previous states, actions, rewards and next states is taken from the memory store and used to update the network, by minimising the Huber loss [32] given by,

$$L = \begin{cases} \frac{1}{2}(x_i - y_i)^2 & for\ |x_i - y_i| \leq 1, \\ |x_i - y_i| - \frac{1}{2}, & otherwise. \end{cases} \quad (4)$$

This loss function acts as a mean squared error loss function when the error falls below 1, and otherwise acts as a simple L1 loss. The Huber loss is used as it is less sensitive to outliers in data than other loss functions such as the mean squared error loss.

After every 10 actions the target network is updated with the current policy networks learned parameters. This ensures that the network is not overfitting and only learning on the most recent data that it has seen, as actions are being chosen using the target network. This is known as off-policy learning, as the actions are chosen by what the target network deems to be give the highest future reward, while we use the actions that the target network decides to update and train the policy network. Here, both networks are made up of the same, 3 hidden layers with 16, 32 and 16 neurones, with 6 input neurones from the design parameters and 12 output neurones for the possible actions, shown in Figure 2. The ADAM optimiser, with a learning rate of 0.001 is used, and the actions are chosen by the Markov decision process using an epsilon greedy policy with a maximum of 0.95, minimum of 0.1 and a decay rate of 0.01. This allows the model to continue to explore with a 10% chance even after a large number of steps, which allows it some chance of escaping any local maxima and continue learning.

The environment is made up of the moth-eye structure on top of a dielectric spacer and metallic reflective layer. Each state represents a geometrically different combination of the structural parameters as shown in Figure 2. P is the periodicity of the hexagonal lattice unit cell, and the diameter of the moth-eye structures, h is the height of the moth-eye structures, c is the radius of curvature, $S_t$ is the thickness of the spacer layer, $S_m$ is the material of the spacer layer and $Sub_m$ is the material of the metallic substrate. The possible materials for the dielectric spacer and metallic reflector are shown in Tables 1 and 2 below. This environment has a total number of possible states of 121 x 121 x 71 x 41 x 9 x 3 = 1,150,738,677 for each metal chosen for the moth-eye structure. It is easy to see that even for a few choices of material, a simple parameter sweep would take a lot of time and computation. Here we found optimised parameters that reach over 90% absorption over the whole range for chromium (Cr), iron (Fe), nickel (Ni), titanium (Ti), vanadium (V) and tungsten (W). Results for other, more traditional metals, gold (Au) and silver (Ag), and a dielectric material, silicon (Si), were also explored and are discussed in the Supplementary Information. Materials such as Cr, Ti, V, and W can be classified as refractory metals, and are resistant to heat and wear, so would be useful in applications such as thermo-photovoltaics. All material data is taken from the Lumerical material database.

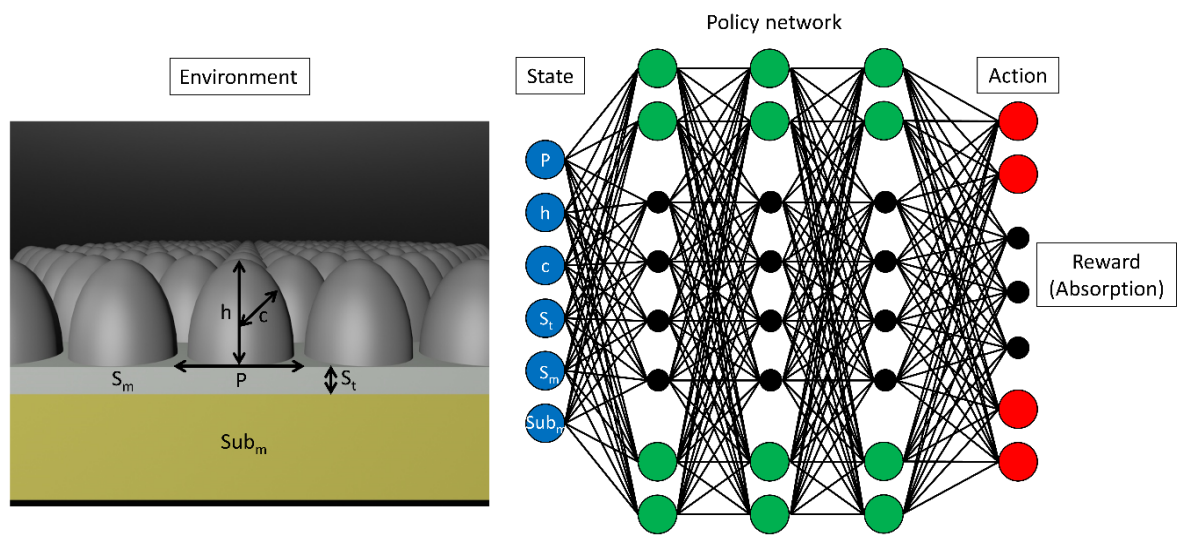

**Figure 2.** Schematic of the moth-eye structure ultra-broadband perfect absorber. The moth-eye structures are made up of a metal, on top of a dielectric spacer layer with a metallic back reflector. The agent, with help from the DDQN, decides the actions to take to achieve the highest possible absorption.

**Table 1.** Possible dielectric materials for the spacer layer.

| Spacer material | |
|---|---|
| **Index** | **Material** |
| 0 | Alumina ($Al_2O_3$) |
| 1 | Silicon (Si) |
| 2 | Silicon Dioxide ($SiO_2$) |

**Table 2.** Possible metallic materials for the metal back reflector.

| Back reflector material | |
|---|---|
| **Index** | **Material** |
| 0 | Aluminium (Al) |
| 1 | Silver (Ag) |
| 2 | Gold (Au) |
| 3 | Chromium (Cr) |
| 4 | Copper (Cu) |
| 5 | Iron (Fe) |
| 6 | Nickel (Ni) |
| 7 | Titanium (T) |
| 8 | Tungsten (W) |

The actions determine what the agent will do to try to maximise its reward and reach its final state as quickly as possible. The actions used here are show below in Table 3.

**Table 3.** The actions available to the agent. The materials are shown in Tables 1 and 2.

| Action Number | Action |
|---|---|
| 0 | Increase the period by 5 nm (maximum 700 nm) |
| 1 | Decrease the period by 5 nm (minimum 100 nm) |
| 2 | Increase the height of the structure by 5 nm (maximum 700 nm) |
| 3 | Decrease the height of the structure by 5 nm (minimum 100 nm) |
| 4 | Increase the curvature by 1 (maximum 70) |
| 5 | Decrease the curvature by 1 (minimum 0) |
| 6 | Increase the thickness of the spacer by 5 nm (maximum 200 nm) |
| 7 | Decrease the thickness of the spacer by 5 nm (minimum 0 nm) |
| 8 | Increase the index of the spacer material by 1 (maximum 2) |
| 9 | Decrease the index of the spacer material by 1 (minimum 0) |
| 10 | Increase the index of the substrate material by 1 (maximum 8) |
| 11 | Decrease the index of the substrate material by 1 (minimum 0) |

After learning about the environment through exploration and exploitation during the training phase, the agent quickly manages to find its way to the optimal parameters to increase the absorption and stop the episode and end with the highest possible reward.

A common problem in reinforcement learning is reward hacking, where the agent learns to repeat the same actions to maximise its reward but without completing the given task as we desire. Such as going around in circles to receive a smaller short term reward many times until the episode ends, rather than taking a suboptimal step now that allows it to reach the terminal state quickly. To overcome this problem, reward shaping was used. If we choose a reward system where the rewards are too sparse, it if difficult for the model to learn, so training would take a long time. So here we chose a reward system that is shaped in a way that low absorption is penalised, and absorption closer to the goal of 99% are greatly rewarded. The reward system was defined as:

$$reward = \begin{cases} -10 & if\ absorption < 85\% \\ (\frac{absorption}{90})^9 - 1 & if\ absorption > 85\% \\ 10,000\ (and\ end\ the\ episode) & if\ absorption > 99\% \end{cases}$$

By using negative rewards, we encourage the agent to finish the episode as quickly as possible as to not keep accumulating negative rewards. The agent starts to get positive rewards only for absorption over 90% and gets a huge positive reward when it reaches the terminal state of 99%. Between 85 and 90%, the rewards are shaped in a way that encourages the agent to choose actions that result in lower negative rewards, i.e. higher absorption, then above 90% absorption the agent starts to receive positive rewards. A visual representation of the reward system is shown below in Figure 3. Three terminal states were included, the first if 99% absorption was reached, the second after 500 steps were taken, and finally if the agent stays in an area where the absorption is less than 85% for over 10 actions. The third terminal state can only be achieved after 150 steps, where the agent will be taking actions from the network ~70% of the time to make sure that exploration isnt hindered. To encourage the agent to end each episode as quickly as possible by reaching the first terminal state, at 99% absorption a large positive reward was given, whereas a large negative reward was given for terminating the episode by the last two methods.

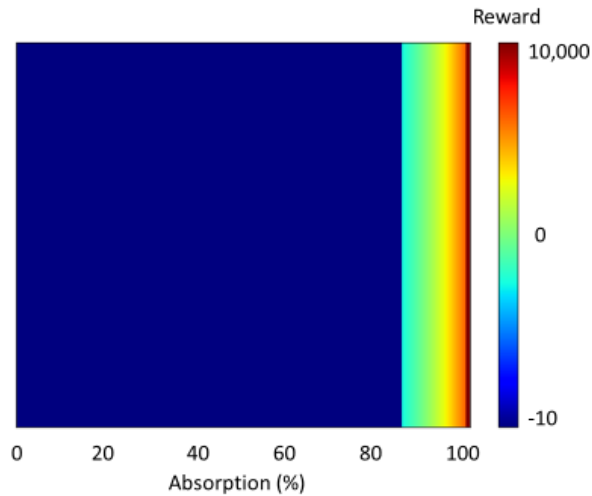

**Figure 3.** A heat map of the rewards available to the agent. The agent receives negative reward between -1 and 0 for absorption over 85% and positive rewards for absorption over 90%, with a large reward of 10,000 for reaching 99% absorption.

After performing 100 episodes of training for the Cr-based perfect absorber, the agent was deemed good enough to be able to test policy in different environments as it was consistently performing well. To test the learned policy, we set the epsilon greedy policy to have a constant value of 0.1, meaning that there is a 10% chance that the model chooses an action randomly, but 90% of the time the agent will use the target network to choose the action, using the knowledge that it learned during the training phase. This is a common procedure in reinforcement learning, as opposed to a fully exploitive model with epsilon of 0, as it allows a small chance of a stochastic element in the policy [33].

Since the initial state of the model is completely random and therefore the number of steps to the optimum state can vary from episode to episode, the number of steps and the cumulative reward per episode are not great indicators of success by themselves. But using the pre-trained policy, the DDQN was able to find parameters for absorption over 90% for in almost every test episode, within around 100-200 steps. This is about $10^7$ times more efficient than a complete parameter sweep for each material. The best results for each material are shown in Figure 4.

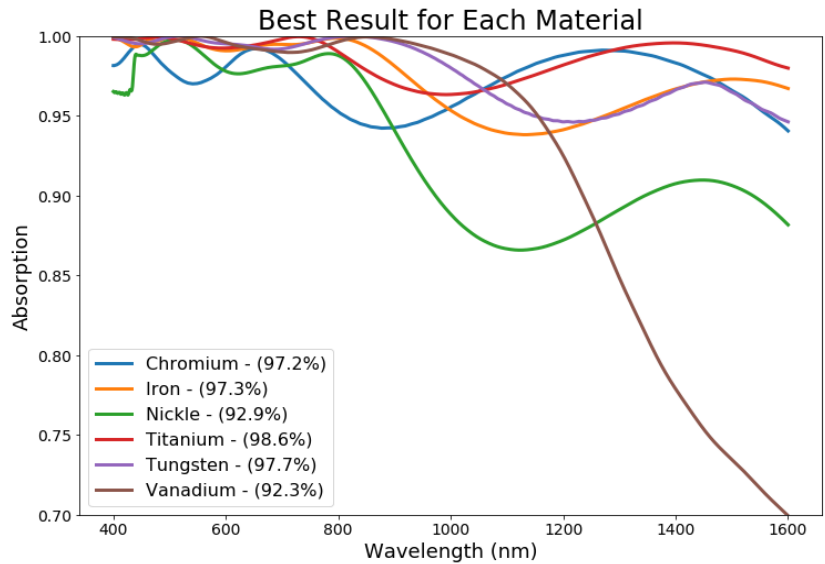

**Figure 4**. Plot of the highest average absorption for each optimised material.

The results for each material are shown in Table 4. Although the materials didn't reach the terminal goal of 99% absorption, the physical limitations of the materials cannot be overcome, so it is highly likely that 99% absorption is not possible. Despite this, designs for absorption over 90% were efficiently found for each material. The imaginary part of the permittivity of all the transition metals used here is very high, over 10, which is an important contributing factor to the absorption.

**Table 4.** The best results of the optimisation process by the DDQN for each moth-eye material.

| Moth-eye Material | Periodicity (nm) | Height (nm) | Radius of Curvature | Spacer Thickness (nm) | Spacer Material | Substrate Material | Average Absorption (%) |
|---|---|---|---|---|---|---|---|
| Chromium | 310 | 620 | 18 | 30 | $SiO_2$ | Ti | 97.2 |
| Iron | 355 | 405 | 37 | 20 | $SiO_2$ | Au | 97.3 |
| Nickel | 505 | 665 | 47 | 100 | $Al_2O_3$ | Cr | 92.9 |
| Titanium | 300 | 515 | 20 | 200 | $Al_2O_3$ | Au | 98.6 |
| Tungsten | 385 | 700 | 25 | 85 | $Al_2O_3$ | Cr | 97.7 |
| Vanadium | 380 | 700 | 24 | 200 | $SiO_2$ | Ni | 92.3 |

To investigate the mechanism of the absorption of the device, we examined the power loss and studied the resulting electric and magnetic field responses at different wavelengths in 100 nm intervals, this can be seen in Figure 5. As expected, the parabolic shape of the moth-eye structure creates a smooth gradient of refractive index, resulting in very low reflection that works in the visible range, where the size of the structures is comparable to the wavelength of visible light. The power is absorbed on the surface of the moth-eye structure while the magnetic field has modes between moth-eye structures and around the tip. For longer, near-IR wavelengths, the electric field is strongly confined to the gaps between the moth-eye structures and the magnetic field becomes strongly confined to between the gaps of the structures. This can be attributed to gap plasmon resonances that can be supported by the highly lossy metals. The high-loss nature of these metals also helps to create the broadband absorber, as the resonances between the back-reflector layer and the moth-eye structures will have a low Q-factor, broadening the absorption further.

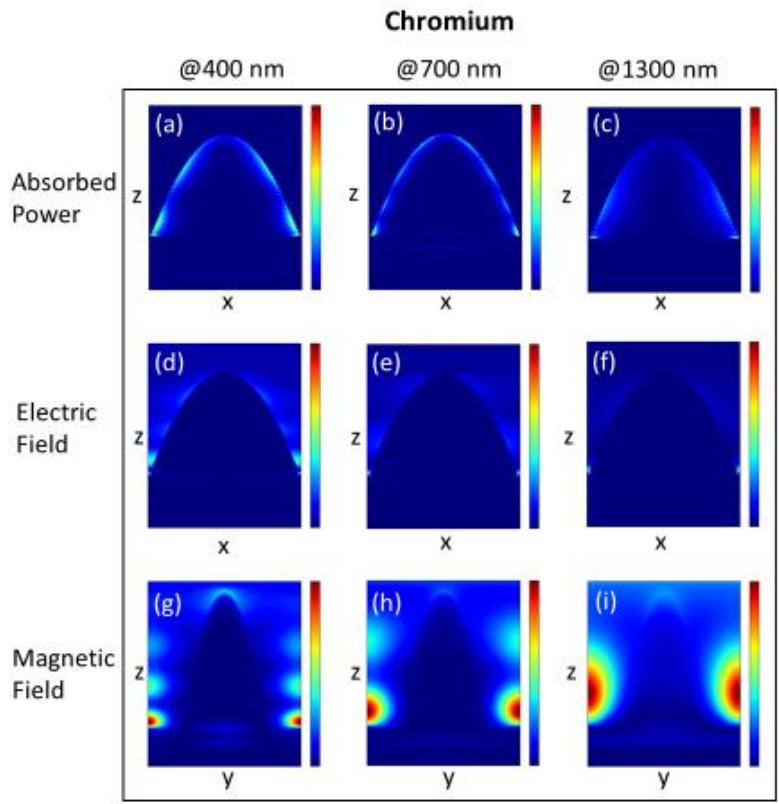

**Figure 5**. (a-c) The absorbed power distribution in the x-z plane for different wavelengths. (d-f) The electric field distribution in the x-z plane and (g-j) magnetic field distributions in the y-z plane. The colour bars represent the minimum (blue) and maximum (red) field.

**Conclusion**

In conclusion, we designed and numerically analysed an ultra-broadband absorber and optimised the performance using a double deep Q-learning algorithm. By training one agent for 100 episodes with a chromium moth-eye structured absorber, we then used the same network to optimise a similar absorber with various other materials to successfully achieve over 90% absorption over the visible and near-IR range usually within ~100-200 steps. This technique could be used for any problems where multiple similar optimisations are needed, and could also possibly be expanded to improve the learning speed for other regimes such as using the knowledge learned in optimising a moth-eye structured absorber, to transfer over to another design.

**Associated Content**

**Supporting Information**

The following files are available free of charge.

Supporting information with discussion of other materials and results not shown in the main text.

**Author Information**

Corresponding author

*Email: jsrho@postech.ac.kr

**Acknowledgements**

J.R. acknowledges the National Research Foundation of Korea (NRF) grants (NRF-2019R1A2C3003129, CAMM-2019M3A6B3030637, NRF-2019R1A5A8080290, NRF-2018M3D1A1058998, NRF-2015R1A5A1037668) funded by the Ministry of Science and ICT (MSIT) of the Korean government. I.K. acknowledges the Global Ph.D. Fellowship (NRF-2016H1A2A1906519) by NRF-MIST of the Korean government.